\documentclass[12pt,letterpaper,fleqn]{article}
\usepackage[utf8]{inputenc}
\usepackage[landscape=false]{geometry}
\usepackage{tabularx}
\usepackage[labelsep=space, labelfont=bf]{caption}
\usepackage{float}
\usepackage{geometry}
\usepackage{xcolor}
\usepackage{array}
\usepackage{longtable}
\usepackage[utf8]{inputenc} %
\usepackage{graphicx} %
\usepackage{threeparttable}%
\usepackage{lipsum}
\usepackage{numprint} 
\usepackage{makecell}
\usepackage{setspace}
\usepackage{amsmath}
\usepackage{soul} %to strike out text and highlight
\setulcolor{yellow} %for highlighting
\usepackage[english]{babel}

\newcommand{\bigchi}{\raisebox{.1\baselineskip}{\large\ensuremath{\chi}}}

\newcolumntype{R}{>{\raggedleft\arraybackslash}p{1.7cm}}
\newcolumntype{C}{>{\centering\arraybackslash}p{0.00cm}}
\newcolumntype{L}{>{\raggedright\arraybackslash}p{3.7cm}}
\newcolumntype{B}{>{\centering\arraybackslash}p{0.00cm}}

\definecolor{nblue}{HTML}{000660}
\PassOptionsToPackage{hyphens}{url}\usepackage[colorlinks=true,urlcolor=nblue,linkcolor=nblue,citecolor=nblue]{hyperref}

\makeatletter
\def\thanks#1{\protected@xdef\@thanks{\@thanks
        \protect\footnotetext{#1}}}
\makeatother

\begin{document}

\renewcommand{\figurename}{Fig.}

\onehalfspacing
\title{\Large \textbf{Modeling Commuter Mobility in Stockholm: A Spatial Panel Approach Using Mobile Phone Data\mbox{}\vspace{0.5cm}}\normalsize}
\author{{Marina Toger\textsuperscript{1}, Umut Türk\textsuperscript{2}, 
    John Östh\textsuperscript{3} and
    Manfred M. Fischer\textsuperscript{4*}}
\thanks{*Corresponding author: Manfred M. Fischer, Vienna University of Economics and Business,\selectlanguage{german} Welthandelsplatz 1,\selectlanguage{english} A-1020 Vienna, Austria; email: \href{mailto:manfred.fischer@wu.ac.at}{manfred.fischer@wu.ac.at}; orcid: 0000-0002-0033-2510\\  
\textsuperscript{1} Department of Human Geography, Uppsala University, Uppsala, Sweden, \textsuperscript{2} Abdullah Gül University, Kayseri, Türkiye, 
\textsuperscript{3} OsloMet University, Oslo, Norway, \textsuperscript{4} Department of SocioEconomics, Vienna University of Economics and Business},  \vspace{-0.5cm}}
\date{}

\vspace{0.5cm}

\maketitle\thispagestyle{empty}\normalsize\vspace*{-2cm}\small

\vspace{3.5cm}
\begin{center}
\begin{minipage}{1\textwidth}
\noindent \textbf{Abstract:} \small

This paper examines the sociodemographic and socioeconomic determinants of regional commuter mobility in the Greater Stockholm Area using a heteroscedastic spatial Durbin panel data model estimated via Bayesian Markov Chain Monte Carlo methods. Drawing on mobile phone–derived origin–destination flows from the MIND database, the analysis exploits unusually fine spatial and temporal granularity across a balanced panel of 675 regions over the period 2018–2023. A $k$-nearest neighbor spatial weight matrix ($k$ = 18), selected via Bayesian model comparison, captures the topological structure of interregional connectivity. By modeling spatial lags in both the dependent and independent variables, the framework enables explicit recovery of direct (own-region) and indirect (spillover) effects from scalar summary measures of the matrix of partial derivatives — providing robust posterior inference on how sociodemographic and socioeconomic conditions propagate through space. This approach addresses a key limitation of conventional non-spatial methods, which risk producing biased estimates by ignoring spatial interdependence. Empirical results confirm that spatial spillovers predominate over direct effects, with educational attainment and car ownership emerging as the principal determinants of commuter mobility, while age composition plays a comparatively modest role. These findings underscore that evaluating direct effects in isolation systematically underestimates the broader societal returns to mobility-enhancing regional policies.

\vspace{5mm}
\end{minipage}

\begin{center}
\normalsize \today{}
\end{center}

\end{center}

\newpage
\section{Introduction} 
Understanding regional commuter mobility is essential to designing effective transportation policies, managing infrastructure demand, and promoting sustainable urban development. Commuter flows are shaped not only by individual and household characteristics, but also by local labor market conditions, accessibility, and spatial interdependence among neighboring regions. Traditional data sources — travel surveys, census records, and administrative commuting statistics — have long formed the empirical backbone of commuter-flow research, yet remain constrained by limited temporal frequency, coarse spatial resolution, and substantial reporting lags (Ortúzar and Willumsen 2011). Consequently, researchers are increasingly using mobile phone data as a valuable proxy to complement or update traditional post-pandemic census datasets, which are often unavailable locally (Martínez-Bernabéu et al. 2025).

During the past decade, mobile phone–based mobility data have transformed the study of commuting and urban travel behavior. Early work demonstrated that call detail records (CDRs)\footnote{Each record registers the user ID, event type, timestamp, and processing base station ID, and by logging the locations of the serving signal towers, these records enable the inference of general human mobility patterns.} and signaling data can reliably reconstruct origin-destination (OD) flows and detect daily and weekly mobility patterns (González et al. 2008; Song et al. 2010). Subsequent studies showed how mobile phone traces can reveal the structure of metropolitan commuting networks (Calabrese et al. 2011; Kung et al. 2014), capture temporal variability in flows (Wang et al. 2018), and inform transportation planning and congestion management (Toole et al. 2015). Further contributions have used mobile phone data to examine polycentricity and urban form (Blondel et al. 2015) and the resilience of commuting systems under disruption (Jiang et al. 2019). Recent methodological developments have even integrated mobile phone tracking grids with local accessibility networks to parse fine-grained intra-city modal shares, such as active transport and public transit (Magyar et al. 2025). This growing literature establishes the value of phone-based mobility data for analyzing fine-grained spatiotemporal dynamics that remain invisible in conventional survey-based datasets.

The present study uses mobile phone–derived OD flows from the MIND database\footnote{MIND denotes the Mobile Individual Network database of the Department of Human Geography at Uppsala University. It contains mobile phone data sourced from a major Swedish mobile network operator, accounting for almost 20 percent of the total Swedish mobile phone market.} to analyze how sociodemographic and socioeconomic conditions shape commuter mobility in the Greater Stockholm Area\footnote{Regions covering the Greater Stockholm Area — comprising Uppsala, Södermanland, and Stockholm counties — correspond to the regional statistical areas defined by Statistics Sweden (2025), referred to throughout this study simply as regions.}. Such highly detailed, individualized spatial big data datasets are crucial for identifying how socioeconomic variables and spatial features interact, as demonstrated in recent space-time accessibility analyses measuring urban inequalities (Barboza et al. 2025). The Greater Stockholm Area represents a metropolitan region characterized by demographic diversity, a polycentric labor market, and strong suburban-urban interactions. These features render spatial dependencies particularly salient: regions are inherently interconnected through economic linkages, transportation infrastructure, and daily movement patterns so that shocks or policy interventions in one area can readily spill over into adjacent ones. Conventional econometric models that ignore such risk of interdependence produce biased or incomplete results (Fischer and LeSage 2025; Elhorst 2014; LeSage and Pace 2009).

This paper contributes to both the commuter mobility and spatial econometrics literature in several respects. First, it applies a heteroscedastic spatial Durbin panel model — still relatively rare in empirical mobility research — to a large high-resolution dataset covering 675 regions over 2018–2023. By modeling spatial lags in both the dependent and independent variables, the framework enables explicit estimation of direct (within-region) and indirect (spillover) effects, which are essential for understanding how sociodemographic and socioeconomic conditions propagate through space. Second, the Bayesian MCMC approach supports full posterior inference, facilitates evaluation of uncertainty in impact estimates, and enables principled comparison of alternative spatial weight specifications (LeSage and Pace 2009). Third, the use of fine-grained phone-based mobility data overcomes the limitations of conventional commuting statistics, offering a more temporally dynamic and spatially detailed portrait of regional mobility in the Greater Stockholm Area.

The remainder of the paper is structured as follows. Section 2 outlines the econometric framework. Section 3 presents the data, the model specification, and the estimation results. The paper concludes with closing remarks.

%================================================================
\section{Econometric Framework}
\subsection{Model specification}
The econometric analysis is based on a spatial Durbin panel data model, which has become a standard specification in the spatial econometrics literature (see, e.g., LeSage 2014; Gopal and Fischer 2023). The model takes the form:
\begin{align}
    y&= \rho W y + X \beta + W X \theta + \iota_T \otimes \mu + \nu \otimes \iota_N + \epsilon.
\end{align}

\noindent The $NT\times 1$ $y$ contains the dependent variable — regional rate of commuter mobility for region $i$ ($i=1,...,N$) at time $t$ ($t=1,..., T$) — stacked with time as slow index. The matrix $NT \times Q$ $X$ contains $Q$ explanatory variables, organized in the same way. The vector $Q \times 1$ $\beta$ collects the associated parameters.

$W$ is the $NT \times NT$ block-diagonal spatial weight matrix $I_T \otimes w$, where $w$ denotes the row-normalized $N\times N$ spatial proximity  matrix with zero diagonal elements. The $NT\times 1$ matrix-vector product $Wy$ is the spatial lag of the dependent variable, reflecting a linear combination of neighboring region values. The scalar parameter $\rho$ measures the degree of spatial dependence in the dependent variable. The $NT\times Q$ matrix product $WX$ generates spatial lags of the $Q$ explanatory variables —  representing linear combinations of characteristics from neighboring regions — with associated parameter vector $\theta$.

The term $\iota_T \otimes \mu$ represents an $N$-vector of region-specific fixed effects $\mu$, repeated for each time period $t$. The term $\nu \otimes \iota_N$ is the Kronecker product of the $T$-vector of time-specific effects $\nu$, one for each period $t$.\footnote{Following LeSage (2021), the region-specific effects $\mu$ and time-specific effects $\nu$ are handled via demeaning transformations rather than assigned explicit priors. For Bayesian inference in spatial Durbin models with heteroscedasticity, see Mills and Parent (2021).}

To accommodate heteroscedastic disturbances, we assume
\begin{align}
           \epsilon  &\sim \mathcal{N}(0_{NT},\sigma^{2} \ V_{NT}),
\end{align}
where $V_{NT}$ is a diagonal variance-covariance matrix with heterogeneous variances $v_{it}$ ($i=1, ..., N$; $t=1, ..., T$) and zero off-diagonal elements, and  $\sigma^2$ denotes the noise variance parameter. The variance scalars $v_{it}$ are treated as unknown parameters to be estimated alongside the remaining model parameters.

%====================================================
\subsection{Bayesian estimation}
The model is estimated using Markov Chain Monte Carlo (MCMC) estimation methods\footnote{MCMC sampling exploits the fact that each full conditional distribution is straightforward to simulate, enabling efficient posterior estimation.}, with prior distributions assigned to the parameters $\beta$, $\theta$, $\sigma$, and $\rho$ as well as the scalar variances $v_{it}$ ($i=1, ..., N$; $t=1, ..., T$). 

For the coefficient vector $\delta = (\beta,\theta)'$, we adopt a joint normal prior:
\begin{align}
    p(\delta) &\sim \mathcal{N}(c, C)
\end{align}
 with prior mean $c$ set equal to one and minimal prior variances (0.001) on the diagonal of the $(2Q)\times(2Q)$ variance-covariance matrix $C$, with zero off-diagonal elements. The tight prior variances bias the estimates toward the prior mean, providing robustness against outliers and non-constant variance across regions and time periods —  a stabilizing property that maximum likelihood estimation cannot deliver (LeSage 2021).

The prior distribution for the variance scalars $v_{it}$  takes the form of a set of $N \times T$ independent and identically distributed ($iid$) chi-square ($r$) distributions\footnote{The use of MCMC estimation with variance scalars of this type was introduced by Geweke (1993). Large posterior estimates for the variance scalars, $v_{it}$, serve to accommodate outliers or observations with high variance, which are consequently down-weighted  —  analogously to generalized least-squares, where large variances yield lower observation weights.}:
\begin{align}
   p(r/{v_{it}}) &\sim  iid \; \bigchi  ^2(r), 
\end{align}
where $r$ is a hyperparameter governing the degree of flexibility in accommodating heteroscedasticity. We set $r=5$ for the chi-squared prior on $v_{it}$, permitting sufficient deviation from the prior mean of one to appropriately adjust for heteroscedasticity in the data. 

A noninformative prior is placed on the noise variance parameter $\sigma^2$, specified as an inverse-gamma distribution:
\begin{align}
p(\sigma^2) &\sim IG (a,b),
\end{align}
where  $a\xrightarrow{}0$, $b\xrightarrow{}0$, reflecting the absence of prior information on this parameter.

Finally, a uniform prior is assigned to the spatial dependence parameter $\rho$: 
\begin{align}
    p(\rho) &\sim U(-1,1)
\end{align}

\noindent constrained to the open interval $(-1,1)$ to ensure invertability of $(I_{NT}-\rho W)$.\footnote{The spatial dependence parameter $\rho$ is restricted to the interval that guaranties the existence of $(I_{NT} - \rho W)^{-1}$. With a  row-normalized $W$, this interval lies between –1 and 1 (LeSage 2021). A lower bound of –1 is imposed to avoid computing eigenvalues of the large inverse matrix.} This constraint is enforced during MCMC estimation via a Metropolis-Hastings approach with rejection sampling, as described in Fischer and LeSage (2020), and LeSage (2020). Following standard practice in the literature, the priors for $\rho,\delta$, and $\sigma^2$ are assumed to be mutually independent.

Estimation proceeds using the conditional distributions outlined in LeSage and Fischer (2020). A Markov chain of 4,000 posterior draws was generated, with the first 500 discarded as burn-in.

%===================================================
\subsection{Direct and indirect impact estimates}
Correct interpretation of spatial regression model estimates requires the use of scalar summary measures, as proposed by LeSage and Pace (2009), who demonstrate that these measures provide the appropriate basis for inference in nonlinear models involving spatial lags of the dependent variable (for an extended discussion, see Fischer and LeSage 2025).

The matrix of partial derivatives $NT\times NT$ used to calculate direct and indirect effect estimates for the $q$th explanatory variable $X^q$ ($q=1, ..., Q$) is given by:
\begin{align}
    \partial y/\partial X^q = (I_{NT}-\hat{\rho} W)^{-1} \ (\hat{\beta} + W \hat{\theta}),
\end{align}
where $\hat{\rho}, \hat{\beta}$ and $\hat{\theta}$ denote posterior parameter estimates. The inverse matrix on the right-hand side is the $NT\times NT$ spatial multiplier matrix that propagates the feedback effects through the spatial network. 

To simplify reporting and interpretation, LeSage and Pace (2009) proposed summarizing the matrix in Eq. (7) through two scalar measures: the mean of the main diagonal elements, yielding a \textit{summary measure of direct effects} (the average impact on a region from a change occurring within that region); and the mean of the row sums of off-diagonal elements, yielding a \textit{summary measure of cumulative indirect effects} (the average cumulative spillover from changes in all other regions). \textit{Total effects} are defined as the sum of direct and indirect effects estimates.

 Inference on these impact estimates is based on simulated draws from the estimated variance-covariance matrix for $\beta$ and $\theta$. A set of 1,000 simulated draws is used to generate empirical distributions for direct, indirect, and total effects. This procedure is implemented efficiently using LeSage's Spatial Econometric Toolbox for MATLAB (LeSage 2021).

%===============================================
\section{Application to the Greater Stockholm Area}

The empirical model is estimated using a balanced pool of $N$=675 contiguous regions covering the Greater Stockholm Area\footnote{The Greater Stockholm Area — comprising Uppsala, Södermanland and Stockholm counties — has a population exceeding three million inhabitants, representing roughly one third of the total Swedish population. At the end of 2023, Sweden had 14.9 million mobile subscriptions, corresponding to approximately 141 subscriptions per 100 inhabitants (Post-och Telestyreslen 2025).} over six years from 2018 to 2023.\footnote{Managing large-scale CDR data is both computationally demanding and time-intensive. To facilitate aggregation from individual trajectories to aggregate movements, the analysis is restricted to a single-day snapshot per year — specifically, the Thursday of week six — capturing one workday's traces of phone user traces. This choice avoids confounding effects from holiday periods, special events, and adverse weather conditions, though the influence of economic cycles and COVID-related disruptions cannot be entirely ruled out.} The dependent variable $y$ represents the regional commuter mobility rates. Ten independent variables believed to influence these rates are included as explanatory factors. All explanatory variables are sourced from Statistics Sweden.\\

%========================================
\subsection{Spatial weight selection}
The spatial lag structure of the model depends on the choice of a spatial weight matrix that summarizes the topological relationships among the regions. A large number of weight matrices can be constructed for any given spatial configuration (Fischer and Wang 2011), but $k$-nearest neighbor matrices — which constrain the neighbor structure to the $k$ closest regions — have gained increasing prominence in spatial econometrics for two main reasons.

First, efficient algorithms are now available in MATLAB for computing the $k$-nearest neighbors on scale, making them practical even for very large datasets. Second, the interpretation of spatial spillovers is more intuitive when every observation unit has the same number of neighbors, as is the case with $k$-nearest neighbor matrices. This contrasts with contiguity-based weight matrices, where the number of neighbors varies across regions, complicating the comparison of spillover magnitudes.

The choice of $k$ is an empirical matter. We resolve it through Bayesian model comparison (LeSage and Fischer 2008), in which model probabilities are derived from the log-marginal likelihood obtained by integrating over the full parameter space. This approach has a notable advantage over likelihood-ratio or Lagrange-multiplier tests: the result does not depend on specific parameter values but holds across the entire parameter space.

Evaluating the $k$-nearest neighbor matrices for $k$ = 1, …, 20 within our spatial panel model specification, we select $k$ = 18 with the highest posterior probability (0.552), indicating that the specification with 18 nearest neighbors is most consistent with the observed panel data. The resulting weight matrix $W$ is a 4,050 × 4,050 block-diagonal matrix, with $w$ representing the spatial proximity matrix normalized in rows, of size 675 × 675, assumed constant over time. The 4,050 × 1 matrix-vector product $Wy$ reflects a linear combination of neighboring region values of the dependent variable, while the 4,050 × 10 matrix product $WX$ generates spatial lags for the ten explanatory variables.

%=================================================
\subsection{Data and data sources}
The mobile phone data are drawn from the MIND database of the Department of Human Geography at Uppsala University. The data aggregation method — a widely used approach in human mobility research (Du et al. 2025) — which summarizes data by tallying the number of trips originating in one region (origin region O) and terminating in another (destination region D).\footnote{This so-called OD aggregation method has been widely adopted in human mobility research (e.g., Iqbal et al. 2014, Alexander et al. 2015).} Accordingly, regional commuter mobility is measured as the sum of regional inflows and outflows, normalized by the resident population.

Although space constraints prevent reporting mobility rates for all 675 regions throughout the full observation window, Figure 1 summarizes them through two geographic maps — one for 2018 and one for 2023. Both maps employ a standard deviation classification scheme and reveal substantial spatial heterogeneity in commuter mobility rates. Rural regions consistently exhibit lower mobility levels than suburban ones, and mobility rates tend to be similar among neighboring regions, providing initial descriptive evidence of spatial dependence.\\

Position FIGURE 1 about here\\

\noindent
Ten explanatory variables are included, selected for their established relevance to regional commuter mobility in the Swedish context.

\textit{Age Composition}. The demographic structure of a region is an important determinant of commuter mobility, as different age groups exhibit distinct travel behaviors shaped by career stage, physical ability, cost sensitivity, and lifestyle preferences. To capture the influence of age structure, three variables are included: the percentage of the population aged 15–24, 25–44, and 45–64 years, respectively.

\textit{Educational Attainment}. Education shapes commuter mobility by influencing job opportunities, income levels, residential choices, and environmental awareness, all of which, in turn, affect commute decisions and travel patterns. Individuals with higher educational qualifications are more likely to hold better-remunerated or more specialized positions distributed across a wider geographic area, making longer commutes a worthwhile trade-off. Educational attainment is measured at the regional level using two variables: the percentage of residents aged 25 and over holding a high school diploma, and the percentage holding a bachelor's degree or higher.

\textit{Foreign Background}. A region's share of foreign-born residents and those born in Sweden to two foreign-born parents is included as a determinant of commuter mobility, given that immigrant populations may face economic constraints limiting car access, exhibit stronger reliance on public transport, and be more strongly influenced by social networks in their commuting decisions — factors that can both encourage and discourage regional mobility.

\textit{Economic Conditions}. Regional economic circumstances shape commuter mobility in complex ways. High unemployment and poverty rates push residents across regional boundaries in search of economic opportunity. In contrast, high income levels expand access to faster and more flexible modes of transport — particularly private cars — and provide greater residential and occupational choice, enabling longer commutes. Three variables are included: the unemployment rate (percentage of residents 20 to 64 years of age who are not employed), the poverty rate (percentage of residents 20 to 64 years of age with income below 60 percent of the national median), and the high income rate (percentage of residents 20 to 64 years of age with income exceeding 200 percent of the national median).

\textit{Car Availability}. The availability of private vehicles directly affects commuter convenience, flexibility, and travel time, particularly in suburban and rural areas where public transport alternatives are limited. Car availability is measured as the number of registered passenger vehicles per 1,000 residents.

%============================================
\subsection{Estimation results}
Table 1 presents posterior means for the parameter estimates of the ten explanatory variables, together with asymptotic $t$-statistics and associated $z$-probabilities. The $t$-statistics are computed from the retained MCMC draws as the ratio of the posterior mean to the posterior standard deviation. The model achieves high goodness-of-fit ($R$-square = 0.84) and a low residual variance (sigma-square = 0.005). The estimated spatial dependence parameter $\hat{\rho}$ = 0.28 is significant at the one percent level, confirming a strong spatial dependence in the dependent variable.

The reliability of the MCMC estimates is assessed using the Geweke (1992) convergence diagnostic, which tests whether the first 10 percent and the last 90 percent of the posterior draws are equal; equality indicates convergence. As reported in Table A of the Appendix, all Geweke $p$-values exceed the conventional 0.05 threshold, confirming that the early and late segments of the MCMC chains do not differ statistically. Monte Carlo errors are negligible relative to posterior standard deviations, and effective sample sizes range from approximately 6,000 to more than 32,000, reflecting adequate chain mixing and high estimation precision. Together, these diagnostics indicate stable convergence and reliable posterior inference.\\

Position TABLE 1 about here\\

Table 2 reports direct, indirect, and total impact estimates together with 95 percent credible intervals, $t$-statistics, and associated probabilities for changes in the $X$- and $WX$-variables.\footnote{Reported posterior means, standard deviations, and 0.95 credible intervals are computed directly from the MCMC draws, and inferences drawn from these correspond to a 95 percent significance level, consistent with maximum likelihood conventions.} A positive (negative) posterior mean with a credible interval lying entirely above (below) zero is interpreted as a statistically significant positive (negative) effect; effects whose credible intervals span zero are not significant. Panel A presents direct impact estimates, Panel B presents indirect impact estimates, and Panel C summarizes total impact estimates.

Turning first to Panel A, seven variables yield positive direct effects, and one yields a negative direct effect that are statistically significant. Educational attainment — at both high school (1.03) and bachelor's degree (1.02) levels — and car availability (0.85) exhibit the largest direct effects. The variables of age composition have moderate positive impacts in the range of 0.2 to 0.3, while the foreign background exerts a small positive effect (0.004). The high income rate is the only variable with a significant negative direct effect ($-$0.003). The direct effects of unemployment and poverty are not significantly different from zero, indicating that these variables do not exert a discernible influence within the region on commuter mobility. It should be noted that direct impact estimates differ from the corresponding coefficient estimates in Table 1 due to feedback effects inherent in the spatial multiplier matrix: where impact estimates exceed coefficients, positive feedback is present; where coefficients exceed impact estimates, negative feedback applies.\\

Position TABLE 2 about here\\                                                                                                                                                                    
\\Panel B presents indirect effects estimates, which measure cumulative spatial spillovers — that is, the aggregate impact of changes in explanatory variables across all other regions $j \neq i$ on mobility outcomes in the region $i$.\footnote{The $\theta$-coefficients do not directly measure spillover magnitudes; indirect effects are obtained from cross-partial derivatives aggregated across all $j \neq i$, as described in Section 2.3.} Indirect effects are consistently larger in magnitude than direct effects, reflecting the cumulative effect of spillovers across all neighboring regions and their neighbors in turn. The spillovers are strongest for educational attainment (high school: 1.69; bachelor's degree: 1.68) and car availability (1.57). Variables of age composition produce smaller but statistically significant spillover effects in the range of 0.11 to 0.14, while poverty generates a modest positive spillover (0.02). The indirect effects of all remaining variables are not significant.

The total impact estimates in Panel C confirm that indirect effects predominate in all significant determinants, underscoring the centrality of spatial spillovers in shaping regional commuter mobility. Inference on total effects is based on the sum of coefficients $\hat{\beta} + \hat{\theta}$, with dispersion estimated from the MCMC draws (Fischer and LeSage 2025). Focusing on direct rather than total impacts would lead to a systematic underestimation of each determinant's influence on commuter mobility across the region.

 %============================================
 \section{Closing Remarks}
This study yields several empirical findings with direct implications for transportation policy and regional planning. First, commuter mobility in the Greater Stockholm Area is shaped not only by the sociodemographic and socioeconomic characteristics of a region itself but also — and more decisively — by those of neighboring regions. Second, spatial spillovers consistently outweigh direct effects: the indirect impact of car availability in neighboring regions, for instance, is nearly twice the magnitude of its direct counterpart. Focusing solely on direct effects, therefore, leads to a systematic underestimation of the societal benefits associated with mobility-enhancing regional policies. Third, educational attainment is the strongest overall determinant of commuter mobility, closely followed by car ownership, while age composition plays a statistically significant but comparatively modest role.

Several limitations merit acknowledgment. The analysis relies on CDR data, which, while analytically powerful, carries an inherent demographic bias: older adults are likely underrepresented among mobile phone users, potentially introducing systematic distortions. Spatial resolution presents a further constraint — inferred locations correspond to signal tower positions, which are densely clustered in urban areas but widely spaced in rural ones, reducing location precision precisely where spatial accuracy matters most. Finally, the scalar summary measures used to characterize direct, indirect, and total effects describe the typical region rather than individual ones, potentially obscuring meaningful spatial heterogeneity in impact estimates. Investigating observation-level variation in these effects would provide a richer empirical basis for spatially differentiated transportation and regional development strategies and would remain a productive avenue for future research.

% ==========================================================
\subsection*{Acknowledgments}
The authors thank James LeSage for generously providing the MATLAB code used in this study, and the Department of Human Geography at Uppsala University for granting access to the MIND database.

% ==========================================================
\subsection*{Authorship Contribution Statement}
All authors contributed to the study conception, interpretation of results, and approval of the final manuscript. \textbf{Manfred M. Fischer} provided general supervision and project guidance, led the methodological development, and drafted, revised, and edited the manuscript. \textbf{Marina Toger} was responsible for data curation and the preparation of Figure 1. \textbf{Umut Türk}  performed the computer simulations, prepared the tables, and established the data repository.

% ==========================================================
\subsection*{Declaration of Competing Interests}
The authors declare no competing financial or non-financial interests relevant to the content of this article.

% ==========================================================
\subsection*{Funding Statement}
This work was supported by Riksbankens Jubileumsfond (RJ) [grant number M24-0047] awarded to Marina Toger and John Östh.

%==============================================
\subsection*{Data Availability Statement}
The balanced panel dataset analyzed in this study covers 675 regions from 2018 to 2023. Regional commuter mobility metrics were derived from mobile phone data sourced from the MIND database (Department of Human Geography, Uppsala University). Explanatory variables were obtained from Statistics Sweden. All supporting data have been deposited in Zenodo and are publicly accessible at https://zenodo.org/records/20181642.

\makeatletter
\newenvironment{mybibliography}
 {\let\@afterindentfalse\@afterindenttrue % we want \parindent anywhere
  \section*{References}%
  \setlength{\leftskip}{\parindent}%
  \setlength{\parindent}{-\leftskip}}
 {}
\makeatother

\begin{mybibliography}

\singlespacing

Alexander L, Jiang S, Murga M, Gonzalez MC (2015) Origin-destination trips by purpose and time of day inferred from mobile phone data, \textit {Transportation Research Part C}, vol. 58, 240-250. doi: 10.1016/j.trc.2015.02.018\\

Barboza M, Giannotti M, Grigolon A, Geurs K (2025) Transit fares and space-time accessibility: Measuring inequalities in park accessibility for São Paulo favela residents using mobile phone and smart card data, \textit {Journal of Transport Geography}, vol. 128, 104395.\\ doi: 10.1016/j.jtrangeo.2025.104395\\

Blondel VD, Decuyper A, Krings G (2015) A survey of results on mobile phone datasets analysis, \textit {EPJ Data Science}, vol. 4:10. doi: 10.1140/epjds/s13688-015-0046-0\\

Calabrese F, Diao M, Di Lorenzo G, Ferreira J, Ratti C (2011) Estimating origin–destination flows using mobile phone location data. \textit {IEEE Pervasive Computing}, vol 10(4), 36–44. doi: 10.1109/MPRV.2011.41\\

Du Y, Aoki T, Fujiwara N (2025) A review of human mobility: Linking data, models, and real-world applications, \textit {Journal of Computational Social Science}, vol. 8:90. doi: 10.1007/s42001-025-00414-7\\

Elhorst JP (2014)  \textit{Spatial Econometrics: From Cross-Sectional Data to Spatial Panels} [Springer Briefs in Regional Science]. Springer, Heidelberg. doi: 10.1007/978-3-642-40340-8\\

Fischer MM, LeSage JP (2020) Network dependence in multi-indexed data on international trade flows, \textit{Journal of Spatial Econometrics}, vol 1(1): article 4, available online. doi: 10.1007/s43071-020-00005-w\\ 

Fischer MM, LeSage JP (2025) \textit{Spatial Econometrics}, for the Encyclopedia of Measurement in Social Sciences, second edition, online published as a stand-alone chapter in the Reference Module in Social Science, Elsevier. Reference Collection in Social Science. doi: 10.1016/B978-0443-26629-4.00074-5\\

Fischer MM, Wang J (2011) \textit{Spatial Data Analysis. Methods and Techniques} [Springer Briefs in Regional Science]. Springer, Heidelberg. doi: 10.1007/978-3-642-21720-3\\

Geweke J (1993) Bayesian treatment of the independent Student t linear model, \textit{Journal of Applied Econometrics}, vol. 8, 19-40. doi: 10.1002/jae.3950080504 \\

Geweke J (1992) Evaluating the accuracy of sampling-based approaches to the calculation of posterior moments. In Bernardo JM, Berger JO, Dawid AP, Smith AFM (eds.) \textit{Bayesian Statistics 4}, pp.169-193. Oxford University Press, Oxford. doi: 10.1093/oso/9780198522669.003.0010 \\

González MC, Hidalgo CA, Barabási A-L (2008) Understanding individual human mobility patterns. \textit{Nature}, vol. 453, 779–782. doi: 10.1038/nature06958\\

Gopal S, Fischer MM (2023) Opioid mortality in the US: Quantifying the direct and indirect impact of sociodemographic and socioeconomic factors, \textit{Letters in Spatial and Resource Sciences}, vol. 16, article 29. doi: 10.1007/s12076-023-00350-y\\

Iqbal MS, Chouhury CF, Wang P, Gonzalez MC (2014) Development of origin-destination matrices using mobile phone call data,  \textit{Transportation Research Part C}, vol. 40, 63-74. doi:  10.1016/j.trc.2014.01.002\\ 

Jiang S, Ferreira J, González MC (2019) TimeGeo: Modeling individual mobility patterns for large-scale travel demand estimation. \textit{Transportation Research Part C}, vol. 108, 275–291. doi: 10.1016/j.trc.2019.10.003\\

Kung KS, Greco K, Sobolevsky S, Ratti C (2014) Exploring universal patterns in human home–work commuting from mobile phone data. \textit{PLOS ONE}, vol. 9(6): e96180. doi: 10.1371/journal.pone.0096180\\

LeSage JP (2014) Spatial econometric panel data model specification: A Bayesian approach, \textit{Spatial Statistics}, vol. 9, 122-145. doi: 10.1016/j.spasta.2014.02.002\\

LeSage JP (2020) Fast MCMC estimation of multiple W-matrix spatial regression models and Metropolis-Hastings Monte Carlo log-marginal likelihoods, \textit{Journal of Geographical Systems}, vol. 22(1), 47-75. doi: 10.1007/s10109-019-00294-2\\

LeSage JP (2021) A panel data toolbox for MATLAB, Department of Economics, University of Toledo\\

LeSage JP, Fischer MM (2008) Spatial growth regressions. Model specification, estimation and interpretation, \textit{Spatial Economic Analysis}, vol. 3(3), 275-304. doi: 10.2139/ssrn.980965\\

LeSage JP, Fischer MM (2020) Cross-sectional dependence specifications in a static trade panel data setting, \textit{Journal of Geographical Systems}, vol. 22(1), 5-46. doi: 10.1007/s10109-019-00298-y\\

LeSage JP, Pace RK (2009) \textit{Introduction to Spatial Econometrics}. CRC Press, Taylor \& Francis Group, Boca Raton. doi: 10.1201/9781420064254 \\

LeSage JP, Pace RK (2021) Interpreting spatial econometric models. In Fischer MM and Nijkamp P (eds.) \textit{Handbook of Regional Science}, 2nd and extended edition, pp. 2201-2218. Springer, Berlin and Heidelberg. doi: 10.1007/978-3-662-60723-7-91\\

Louail T, Lenormand M, Picornell M, Garcia Cantu O, Herranz R, Frias-Martinez E, Ramasco JJ, Barthelemy M (2015) Uncovering the spatial structure of mobility networks, \textit{Nature Communications},  6: 6007. doi: 10.1038/ncomms7007\\

Louail T, Lenormand M, Picornell M, Garcia Cantu O, Herranz R, Frias-Martinez E, Ramasco JJ, Barthelemy M (2015) Uncovering the spatial structure of mobility networks, \textit{Nature Communications},  6: 6007. doi: 10.1038/ncomms7007\\

Magyar M, Ala-Hulkko T, Antikainen H, Lankila T, Kotavaara O (2025) Utilizing mobile phone tracking data to estimate intra-city modal mobility: A study on active mobility in two Finnish city regions, \textit {Journal of Transport Geography}, vol. 128, 104326. doi: 10.1016/j.jtrangeo.2025.104326\\

Martínez-Bernabéu L, Coombes M, Casado-Díaz JM (2025) Assessing mobile phone data as proxy census commuting data for transport geography research: A critical review and case study, \textit {Journal of Transport Geography}, vol. 128, 104361. doi: 10.1016/j.jtrangeo.2025.104361\\

Mills JA, Parent O (2021) Bayesian Markov Chain Monte Carlo estimation. In Fischer MM and Nijkamp P (eds.) \textit{Handbook of Regional Science}, 2nd and extended edition, pp. 2073-2096. Springer, Berlin and Heidelberg. doi: 10.1007/978-3-662-60723-7-89\\

Ortúzar J de D, Willumsen L (2011) \textit{Modelling Transport}, 4th edition. Wiley, Chichester. doi: 10.1002/9781119993308\\

Post- och Telestyrelsen (2025) \textit{Svensk telekommarknad 2024} (PTS-ER-2025: 9). \\https://lnk.ua/BNEM7YDNG\\

Song C, Qu Z, Blumm N, Barabási A-L (2010) Limits of predictability in human mobility. \textit{Science}, vol. 327, 1018–1021. doi: 10.1126/science.1177170\\

Statistics Sweden (2025) Regional Statistical Areas (RegSO), 2018 version [Spatial Dataset] Stockholm: Statistics Sweden. Retrieved from [https://lnk.ua/k4kjAd8NL]\\

Toole JL, Colak S, Sturt B, Alexander LP, Evsukoff A, González MC (2015) The path most traveled: Travel demand estimation using big data resources. \textit{Proceedings of the National Academy of Sciences}, vol. 112(3), 881–886. doi: 10.1073/pnas.1404364112\\

Wang Q, He Z, Leung Y, Tian Y (2018) Temporal variability in commuting flows: A multiscale perspective using mobile phone data. \textit{Transportation Research Part A}, vol. 111, 44–60. doi: 10.1016/j.tra.2018.03.018\\

\end{mybibliography}

\onehalfspacing
\clearpage
\noindent
\section*{Appendix}

% Reset table counter and change format for appendix
\setcounter{table}{0}
\renewcommand{\thetable}{A}

\begin{table}[ht]
\centering
\captionsetup{singlelinecheck = false, format= hang, justification=raggedright, labelsep=space, width=1.03\textwidth} 
\caption{MCMC convergence diagnostics for  the spatial Durbin panel data model}
\footnotesize
\begin{tabular}{lrrrrrrr}
\hline
Parameter & Mean & StdDev & MC Error & Tau & ESS & Geweke Z & Geweke $p$ \\
\hline
Age 15--24 ($\beta_{1}$) & 0.0247 & 0.0038 & 0.0000 & 1.42 & 21154.3 & $-$0.395 & 0.307 \\
Age 25--44 ($\beta_{2}$) & 0.0303 & 0.0040 & 0.0000 & 1.87 & 16015.0 & $-$0.389 & 0.303 \\
Age 45--64 ($\beta_{3}$) & 0.0202 & 0.0040 & 0.0000 & 1.95 & 15356.4 & $-$2.281 & 0.977 \\
High School ($\beta_{4}$) & 0.9770 & 0.0315 & 0.0002 & 0.91 & 32807.0 & $-$1.342 & 0.821 \\
Bachelor ($\beta_{5}$) & 0.9648 & 0.0311 & 0.0002 & 1.24 & 24278.2 & 1.806 & 0.929 \\
Foreign Background ($\beta_{6}$) & 0.0040 & 0.0012 & 0.0000 & 1.85 & 16259.5 & $-$0.846 & 0.603 \\
Unemployment ($\beta_{7}$) & $-$0.0014 & 0.0014 & 0.0000 & 1.64 & 18242.7 & 0.237 & 0.188 \\
Poverty ($\beta_{8}$) & 0.0028 & 0.0016 & 0.0000 & 1.88 & 15990.4 & 0.735 & 0.538 \\
High Income ($\beta_{9}$) & $-$0.0031 & 0.0015 & 0.0000 & 1.45 & 20717.9 & $-$1.387 & 0.835 \\
Car Availability ($\beta_{10}$) & 0.8158 & 0.0303 & 0.0002 & 1.38 & 21805.7 & 1.305 & 0.808 \\
W × Age 15--24 ($\theta_{1}$) & 0.1762 & 0.0139 & 0.0001 & 2.19 & 13688.4 & 0.650 & 0.484 \\
W × Age 25--44 ($\theta_{2}$) & 0.3762 & 0.0177 & 0.0002 & 2.94 & 10188.4 & $-$0.416 & 0.323 \\
W × Age 45--64 ($\theta_{3}$) & 0.3108 & 0.0163 & 0.0001 & 2.44 & 12314.9 & $-$0.087 & 0.069 \\
W × High School ($\theta_{4}$) & 0.9956 & 0.0317 & 0.0002 & 1.33 & 22473.6 & 1.491 & 0.864 \\
W × Bachelor ($\theta_{5}$) & 0.9906 & 0.0315 & 0.0002 & 0.94 & 31844.5 & $-$0.362 & 0.283 \\
W × Foreign Background ($\theta_{6}$) & $-$0.0150 & 0.0044 & 0.0000 & 1.58 & 18946.3 & 1.502 & 0.867 \\
W × Unemployment ($\theta_{7}$) & $-$0.0060 & 0.0050 & 0.0000 & 1.35 & 22148.3 & 0.587 & 0.442 \\
W × Poverty ($\theta_{8}$) & 0.0376 & 0.0061 & 0.0000 & 1.53 & 19565.7 & $-$1.623 & 0.895 \\
W × High Income ($\theta_{9}$) & $-$0.0023 & 0.0044 & 0.0000 & 1.35 & 22146.9 & 0.974 & 0.670 \\
W × Car Availability ($\theta_{10}$) & 0.9938 & 0.0312 & 0.0002 & 0.94 & 31875.2 & 0.971 & 0.668 \\
Rho ($\rho$) & 0.5418 & 0.0224 & 0.0002 & 1.43 & 20917.0 & $-$1.100 & 0.729 \\
Sigma-Square & 0.0060 & 0.0002 & 0.0000 & 4.88 & 6144.0 & 0.275 & 0.217 \\
\hline
\end{tabular}
\vspace{2mm}\\

{\raggedright \footnotesize\textit{Notes}: ESS denotes the effective sample size. The Geweke statistic tests for equality of means between the first 10 percent and the last 90 percent of the MCMC draws; equality indicates convergence. All Geweke \textit{p}-values exceed the conventional 0.05 threshold, confirming that the early and late segments of the MCMC chains do not differ statistically. Monte Carlo error values are negligible relative to posterior standard deviations, indicating high estimation precision. Effective sample sizes range from approximately 6,000 to over 32,000, reflecting adequate chain mixing and sufficient posterior draws. Taken together, the diagnostics indicate stable convergence and reliable posterior inference. \par}

\end{table}

\newpage
\section*{Figure to be included in Section  3.2}
% Reset table counter and change format for appendix
\setcounter{table}{0}
\renewcommand{\thetable}{\arabic{table}}

\begin{figure}[htp]
    \centering
    \includegraphics[width=\linewidth]{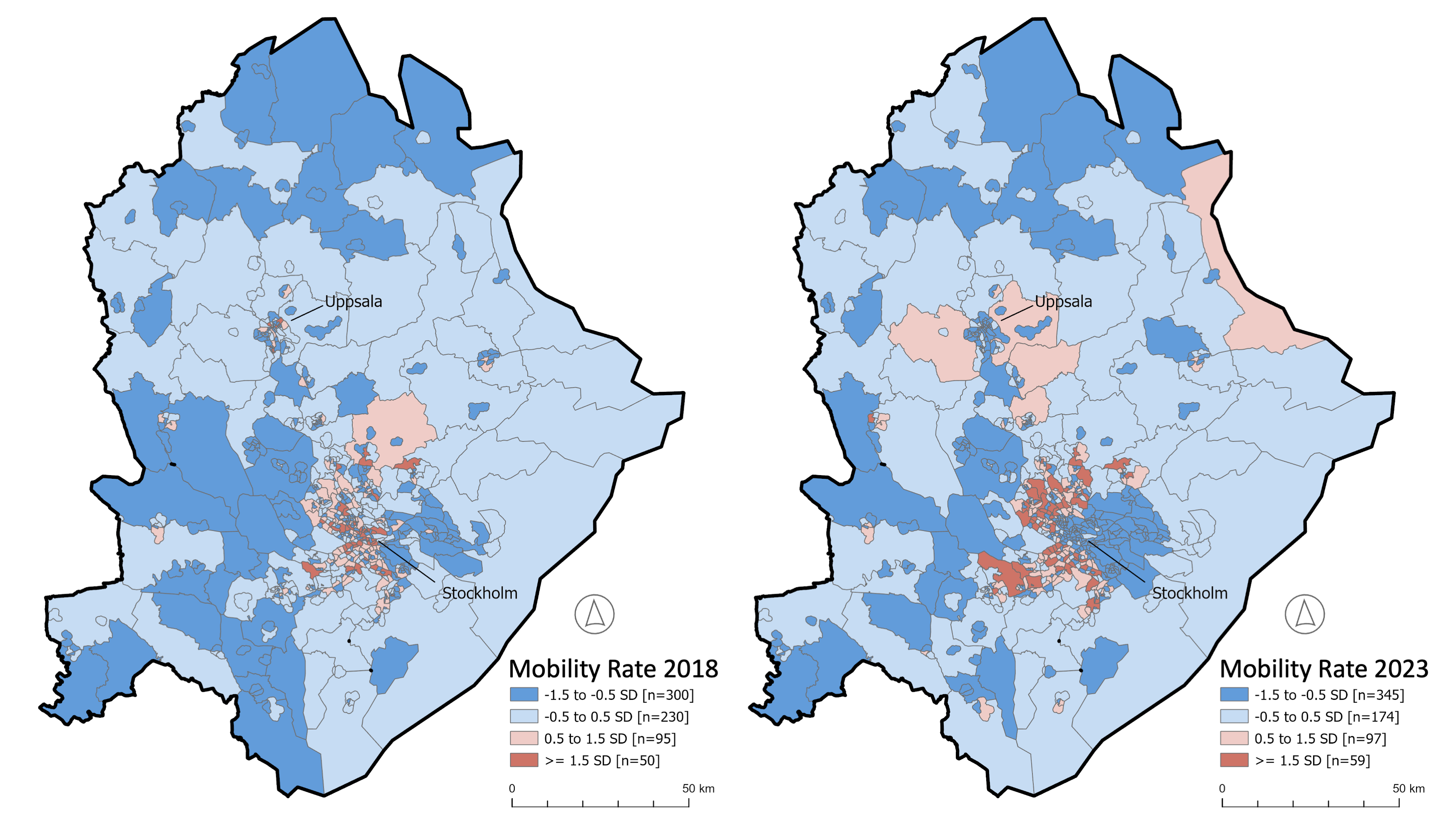}
    \caption{Regional commuter mobility rates in the Greater Stockholm Area, 2018 and 2023, measured as the sum of regional inflows and outflows normalized by the resident population. Class breaks are generated using a standard deviation classification scheme, applied above and below the mean mobility rate of 0.211 (2018) and 0.198 (2023), with standard deviations of 0.300 (2018) and 0.291 (2023), respectively. The number of regions assigned to each class is indicated in square brackets. Thin lines denote the boundaries of the 675 individual regions; thick lines delineate the outer boundaries of the Greater Stockholm Area.} 
    
    \label{fig:commobility}
     
\end{figure}

\newpage
\section*{Tables to be included in Section  3.3}
% Reset table counter and change format for appendix
\setcounter{table}{0}
\renewcommand{\thetable}{\arabic{table}}

\begin{footnotesize}
\begin{table}[H]
    %\centering
    \setstretch{1.17}
    \captionsetup{font=normalsize, width=0.99\textwidth, justification=raggedright, singlelinecheck=false}
    \label{tab:post_est}
    \caption{Posterior parameter estimates of the spatial Durbin panel data model}
    \footnotesize
    \begin{tabular}{l rrr rrr rrrr}
    \cline{1-7}\\[-0.5em]
    Variable (Parameter)      &  &\begin{tabular}[r]{@{}r@{}}Posterior\\Mean\end{tabular} &  &\begin{tabular}[r]{@{}r@{}}Asymp.\\\textit{t}-stat.\end{tabular}   &  & \textit{z}-prob. &  & &  &  \\ \cline{1-7}\\[-0.5em]
    Age 15-24($\beta_1$)        &  &  0.0218 &   &   6.0340 & &        0.0000  &  &            &  & \\[0.5pt]
    Age 25-44 ($\beta_2$)        &  &      0.0244  & &       6.5674 & &      0.0000      &  &            &  &            \\[0.5pt]
    Age 45-64 ($\beta_3$)       &  &    0.0158    & &   4.3296  & &       0.0000     &  &            &  &            \\[0.5pt]
    High School ($\beta_4$)       &  &  0.9739 
    & &      31.0619    & &     0.0000  &  &            &  &            \\[0.5pt]
    Bachelor ($\beta_5$)     &  &    0.9586    & &     30.7538 & &       0.0000     &  &            &  &            \\[0.5pt]
    Foreign Background ($\beta_6$)      &  &  0.0040     & &   3.4457  & &      0.0006  &  &            &  &            \\[0.5pt]
    Unemployment ($\beta_7$)       &  &    $-$0.0001   & &      $-$0.0705    & &    0.9438  &  &            &  &            \\[0.5pt]
    Poverty ($\beta_8$)        &  &     0.0010     & &    0.6756    & &    0.4993 &  &            &  &            \\[0.5pt]
    High Income ($\beta_{9}$)        &  &   $-$0.0031    & &     $-$2.2324    & & 0.0256  &  &            &  &            \\[0.5pt]
    Car Availability($\beta_{10}$)          &  &  0.8006    & &   26.5672   &  &     0.0000    &  &            &  &            \\[0.5pt]    
    $W*$ Age 15-24 ($\theta_{1}$)         &  &    0.0762  & &          8.7582 & &         0.0000    &  &            &  &            \\[0.5pt]
    $W*$ Age 25-44 ($\theta_{2}$)          &  &    0.1248   & &        12.9910 & &         0.0000   &  &            &  &            \\[0.5pt]
    $W*$ Age 45-64 ($\theta_{3}$)        &  &    0.0996   & &       10.8347  & &      0.0000    &  &            &  &            \\[0.5pt]
    $W*$ High School ($\theta_{4}$)         &  &     0.9927   & &        31.3261  & &         0.0000   &  &            &  &            \\[0.5pt]
    $W*$ Bachelor ($\theta_{5}$)      &  &      0.9860    & &         31.2280 & &        0.0000    &  &            &  &            \\[0.5pt]
    $W*$ Foreign Background ($\theta_{6}$)        &  &     $-$0.0051    & &      $-$1.8531   & &     0.0639      &  &            &  &            \\[0.5pt]
    $W*$ Unemployment ($\theta_{7}$)        &  &  0.0020  & &        0.7117 & &         0.4766      &  &            &  &            \\[0.5pt]
    $W*$ Poverty ($\theta_{8}$)        &  &    0.0123   & &      3.6855   & &      0.0002       &  &            &  &            \\[0.5pt]
    $W*$ High Income  ($\theta_{9}$)       &  &   $-$0.0016    & &     $-$0.5748    & &     0.5654   &  &            &  &            \\[0.5pt]
    $W*$ Car Availability ($\theta_{10}$)          &  &     0.9506   & &     30.8732   & &      0.0000    &  &            &  &            \\[0.5pt]
    Rho ($\rho$)  &  &      0.2773 & &         14.1188    & &         0.0000      &  &          & &         \\[0.5pt]
\\\cline{1-7}\\[-1.7em]
\end{tabular}
\vspace{5mm}\\
{\raggedright \footnotesize\textit{Notes}: Estimation is based on a \textit{k}-nearest neighbor spatial weight matrix $W$ with \linebreak $k=18$ nearest neighboring regions. The panel comprises  $N=675$ regions and  $T=6$ \linebreak time periods, with $Q=10$ explanatory variables. Parameter estimates are produced \linebreak using LeSage's (2021) Panel Data Toolbox for MATLAB. Model fit statistics: \linebreak log-marginal likelihood=$-$2643.03, $R$-square=0.8376, and sigma-square=0.0052. \par}
\end{table}
\end{footnotesize}

\newpage
{\setstretch{1.25}
\footnotesize
\begin{longtable}{lrrrrr}%{p{2.5cm}p{0.5cm}p{2cm}%p{4cm}p{3.5cm}p{2.5cm}}%{LBRCRCRCRCR}
%\toprule 
%\hline\\[-0.5em]
\captionsetup{singlelinecheck = false, format= hang, justification=raggedright, labelsep=space, width=1.03\textwidth} %font=footnotesize, 
\caption{Direct, indirect, and total impact estimates of sociodemographic and socioeconomic determinants of regional commuter mobility}

%  Variable      &  & Lower 0.01 &  & Lower 0.05 &  & Median &  & Upper 0.95 &  & Upper 0.99
%\midrule
\endfirsthead 
%\bottomrule
\hline\\[-0.5em] \endfoot 
%\multicolumn{6}{r}{\textit{ctd.}}\\\toprule 
\hline\\[-1em]
 Impact Estimate      &  Posterior Mean & \textit{t}-stat. & \textit{t}-prob. & Lower 0.05  &  Upper 0.95\\[1em]
 \hline %[-0.5em]
%\midrule \endhead 
%\bottomrule \endlastfoot 
%{\large Subtable B}\\[10pt]
%\hline\\[-1em]
%Impact Estimates     &  Posterior Mean  & \textit{t}-stat. & \textit{t}-prob. & Lower 0.05 &  Upper 0.95 \\[1em] \hline\\[-1em]
\textit{A. Direct Impact Estimates} &        &        &  &            &    \\[4pt]
% \textit{}     &  &            &  &            &  &        &  &            &  &            \\[5pt]
Age 15-24        &       0.0255    &       6.8610     &  0.0000   &       0.0183     &      0.0329\\[0.5pt]
Age 25-44       &       0.0304      &    7.8178   &      0.0000     &    0.0229     &     0.0381 \\[0.5pt]
Age 45-64      &      0.0206    &     5.4707     &     0.0000    &    0.0133   &   0.0281\\[0.5pt]
High School        &        1.0319       &    31.9839      &   0.0000    &     0.9678     &     1.0949 \\[0.5pt]
Bachelor    &        1.0161   &        31.6368    &      0.0000     &     0.9539  &        1.0792 \\[0.5pt]
Foreign Background       &     0.0038     &     3.2595 &        0.0011 &       0.0015 &          0.0061            \\[0.5pt]
Unemployment        &      0.0000  &          0.0006 &        0.9995 &        $-$0.0025 &          0.0025           \\[0.5pt]
Poverty        &     0.0016      &     1.0605 &          0.2890 &         $-$0.0013 &          0.0044        \\[0.5pt]
High Income        &     $-$0.0032  &         $-$2.3175 &         0.0205 &         $-$0.0059 &          $-$0.0005       \\[0.5pt]
Car Availability        &     0.8545    &     27.5606 &          0.0000 &         0.7932 &         0.9151      \\[4pt]
\hline
\textit{B. Indirect Impact Estimates} &             &             &            &            &              \\[4pt]
 %\textit{ }    &  &            &  &            &  &        &  &            &  &            \\[5pt]
Age 15-24       &          0.1101    &      8.8615   &      0.0000     &     0.0864     &     0.1349 \\[0.5pt]
Age 25-44        &        0.1762    &       12.2189   &        0.0000     &     0.1487    &       0.2054 \\[0.5pt]
Age 45-64     &         0.1393       &    10.5187       &    0.0000       &   0.1138     &      0.1658 \\[0.5pt]
High School       &          1.6913  &        20.8133  &        0.0000     &    1.5369    &      1.8544 \\[0.5pt]
Bachelor    &           1.6766    &       20.8139    &       0.0000     &     1.5235   &        1.8396 \\[0.5pt]
Foreign Background          &        $-$0.0054     &       $-$1.4596   &        0.1445    &     $-$0.0128  &        0.0018 \\[0.5pt]
Unemployment        &        0.0027    &       0.7031     &      0.4820   &      $-$0.0047    &       0.0101 \\[0.5pt]
Poverty       &          0.0168   &      3.7433    &       0.0002   &       0.0081    &      0.0257 \\[0.5pt]
High Income       &         $-$0.0033    &       $-$0.8982   &       0.3691   &       $-$0.0104   &        0.0038 \\[0.5pt]
Car Availability        &         1.5704    &      21.1112    &       0.0000    &      1.4294   &     1.7210 \\[4pt]
\hline
\textit{C. Total Impact Estimates} &             &            &         &             &              \\[4pt]
 %\textit{}     &  &            &  &            &  &        &  &            &  &            \\[5pt
%&           0.2767*   &         2.2097     &       0.0276    &        0.0392    &      0.5319  \\[0.5pt]
Age 15-24         &         0.1357    &       9.7027    &      0.0000     &     0.1090   &      0.1636  \\[0.5pt]
Age 25-44       &         0.2066  &       12.7091   &         0.0000      &     0.1756    &      0.2395  \\[0.5pt]
Age 45-64          &        0.1599     &      10.8803     &       0.0000    &       0.1317   &        0.1891  \\[0.5pt]
High School     &          2.7232  &        28.3041    &       0.0000   &        2.5340    &       2.9165  \\[0.5pt]
Bachelor        &           2.6928     &     28.1505  &         0.0000    &       2.5112   &        2.8850  \\[0.5pt]
Foreign Background        &           $-$0.0016   &      $-$0.3977   &        0.6909    &      $-$0.0095  &         0.0063  \\[0.5pt]
Unemployment       &           0.0027   &       0.6361   &        0.5248    &     $-$0.0054  &       0.0109  \\[0.5pt]
Poverty      &            0.0184   &       3.7164   &        0.0002   &      0.0088 &         0.0281 \\[0.5pt]
High Income        &          $-$0.0065  &       $-$1.6435   &        0.1004    &     $-$0.0141  &        0.0013  \\[0.5pt]
Car Availability   &      2.4249   &     27.2020   &      0.0000   &    2.2562 &       2.6066  \\[0.5pt]
\end{longtable}}
%{ \footnotesize \setstretch{1}
%\end{tabular}
%\vspace{1mm}\\

{\raggedright \footnotesize\textit{Notes}: Panel A reports direct impact estimates (average within-region effects); Panel B reports indirect impact estimates (average cumulative spillover effects across neighboring regions); Panel C reports total impact estimates, defined as the sum of direct and indirect effects. A positive (negative) posterior mean with a credible interval lying entirely above (below) zero should be interpreted as a statistically significant positive (negative) effect; effects whose credible intervals span zero are not significant. All impact estimates are computed using LeSage's (2021) Panel Data Toolbox for MATLAB.\par}
%\end{table}
%\end{footnotesize}
%________________________________________

\vspace{3mm}

\end{document}